# Interaction Dynamics of Borrelia Surface Proteins with Fibronectin

Kavindi Madduma Hewage[1], Carlos Munoz[1], Mehmet Ozdogan[1], Catherine A. Brissette[2]*, Nuri Oncel[1]*

1 Department of Physics and Astrophysics, University of North Dakota, Grand Forks, ND, 58202, USA

2 Department of Biomedical Sciences, University of North Dakota, Grand Forks, ND, 58202, USA

* Correspondence should be directed to catherine.brissette@und.edu & nuri.oncel@und.edu

**Abstract**

Lyme disease, caused by the bacterium *Borrelia burgdorferi*, is a significant public health concern in North America, with approximately 500,000 cases reported annually in the United States. The dissemination of *B. burgdorferi* from the initial tick bite site to various tissues is facilitated by surface adhesins that bind to extracellular matrix (ECM) proteins such as fibronectin (Fn). This study investigates the binding dynamics of *B. burgdorferi* surface proteins RevA, BBK32, BmpA, OspA, and OspC to Fn using atomic force microscopy-based single-molecule force spectroscopy (AFM-SMFS). Our results demonstrate that RevA and BBK32 form strong, stable bonds with Fn, highlighting their roles as key mediators of host-cell attachment. By quantifying the rupture forces and kinetic parameters of these interactions, we provide a deeper understanding of *B. burgdorferi* adhesion mechanics and offer insights into potential therapeutic strategies targeting early bacterial attachment.

**Statement of Significance**

Lyme disease affects hundreds of thousands of people in the U.S. each year, yet how the bacteria responsible spread through the body remains poorly understood. This study uses a highly sensitive technique—single-molecule force spectroscopy—to measure how bacterial surface proteins attach to human tissue proteins. The results reveal that two proteins, RevA and BBK32, form especially strong bonds with fibronectin, a key structural component in human tissues. Understanding these molecular interactions provides new insights into how the bacteria establish infection and may help guide the development of therapies that block early bacterial attachment, potentially reducing disease progression. This work bridges microbiology and biophysics to address a pressing public health issue.

**Introduction**

Lyme disease is the most frequently reported vector-borne illness in North America, with approximately 500,000 cases in the U.S. each year—a number that continues to climb as *Ixodes* ticks carrying *Borrelia burgdorferi* expand their range. [1–3] If untreated, the infection can disseminate from the bite location to joints, skin, the nervous system, and the heart, [2] and a subset of patients suffer persistent, debilitating symptoms that pose a substantial public-health burden. [4] Dissemination depends on a broad range of *B. burgdorferi* surface adhesins that bind proteins in the extracellular-matrix (ECM) such as decorin, fibronectin (Fn), and laminin, enabling colonization of diverse tissues. [5–8] Blocking these early adhesive interactions is therefore an attractive route to disease prevention. Among the known adhesins, DbpA/B, BBK32, RevA, and

OspC are regarded as key mediators of tissue attachment despite their overlapping binding specificities [8], yet the dynamics of their bonds remain only partially quantified.[9,10]

Table 1: List of proteins used in this study.

| Protein | Function | References |
|---|---|---|
| OspA | Outer surface protein (Osp) | [11,12] |
| OspC | Multi-functional (binds plasminogen, complement) major outer surface protein | [13–16] |
| FlaB | Flagellin: internal protein control | [17] |
| RevA | Fn and laminin binding protein | [18,19] |
| BBK32 | Fn, GAG, complement-binding protein | [20–24] |
| BmpA | Laminin binding protein | [25] |

Developments in the fields of microscopy have provided greater insight into the biology and mechanisms of infection; however, current optical techniques have great difficulty in directly measuring protein-protein interactions that are key to pathogenicity. With the development of probe-based techniques, such as atomic force microscopy, *in situ*-like measurements can be made at the nanoscale, providing insight into the dynamical mechanisms behind molecular binding. Using an extension of AFM known as AFM-single molecule force spectroscopy (AFM-SMFS), cantilever probes are functionalized with ligands/receptors that can be used to directly measure binding affinities and kinetic properties of molecular systems. Under the right conditions, AFM-SMFS can measure individual ligand-receptor interactions in a near physiological regime by pulling on single molecules and recording extension forces with piconewton sensitivity. [9]

Using AFM-SMFS, we probed how *B. burgdorferi* surface proteins bind to fibronectin, a prominent ECM glycoprotein. Fn-functionalized AFM tips [18] were brought into contact with substrates coated with RevA, BBK32, BmpA, OspA, or OspC, while FlaB (flagellin) and Bovine Serum Albumin (BSA, as negative controls). Our measurements show that RevA and BBK32 form strong, stable bonds with fibronectin, highlighting these adhesins as prime drivers of host-cell attachment. [18] By quantifying their rupture forces and kinetic parameters, we improve the current understanding of *B. burgdorferi* adhesion mechanics and provide a foundation for therapeutic strategies that target early bacterial attachment.

**Experimental**

*Protein Preparation*

OspA, OspC, and FlaB recombinant proteins were purchased from Rockland Immunochemicals, Limerick, PA; 000-001-C13 (OspA), 000-001-C11 (OspC), 000-001-C14 (FlaB). Recombinant BmpA was purchased from Ray Biotech, Peachtree Corners, GA; 228-11811. Correct protein sizes, folding, and recognition by specific antibodies were determined by the manufacturers. Recombinant RevA and BBK32 were produced in our laboratory. DNA sequences were amplified via PCR using DeepVent DNA Polymerase (NEB, #M0258S; Ipswich, MA). The nucleotide sequences were then cloned into the pET200 expression vector (Life Technologies, #K200-01), transformed into TOP10 E. coli, and the cells were plated on LB agar supplemented with kanamycin (50 mg/mL; MP Biomedicals, # 0215002925; Santa Ana, CA). The resulting colonies were screened by colony PCR for the appropriately sized insert and further verified by bidirectional sequencing of purified plasmid DNA, as we have previously described.[18,25]

To produce proteins, plasmids with appropriately inserted coding regions were transformed into Rosetta (DE3) pLysS E.coli and plated on LB agar with kanamycin (50 mg/mL) and chloramphenicol (30 mg/mL–Sigma, #C-0378). Individual colonies were selected and inoculated into super broth (SB–32 g Tryptone, 20 g yeast extract, 10 g NaCl, per liter of $H_2O$) starter culture overnight, supplemented with kanamycin and chloramphenicol as above. The SB starter culture was transferred (1:50 dilution) into the final cultures of antibiotic-supplemented SB and allowed to grow to an OD of 0.5, then induced with 0.1–0.3 mM isopropyl β-D-1-thiogalactopyranoside (IPTG) for four hours. Cells were spun down, re-suspended in MagneHis wash buffer (MHWB-100 mM HEPES, 10 mM imidazole), and lysed via sonication with a Model 705 Sonic Dismembrator (Fisher Scientific; Pittsburgh, PA). The recombinant proteins were purified with Nickel bead affinity chromatography using MagneHis Ni-Particles (Promega, #V8565; Madison, WI) and magnetic stands. Four washes using MHWB were used to remove contaminating proteins, and MagneHis elution buffer (100 mM HEPES, 500 mM imidazole) was used to elute the purified recombinant proteins from the Nickel particles. This final elution buffer was replaced with phosphate-buffered saline via dialysis in 3,500 MWCO dialysis cartridges (Pierce, #66330; Rockford, IL). Final protein concentrations were determined by a BCA assay (Pierce, #23227), and purities were assessed by SDS-PAGE, which was then stained with Coomassie brilliant blue dye. Recombinant RevA and BBK32 produced in this manner bind fibronectin in enzyme-linked immunosorbent assays (ELISA) and are correctly recognized by specific antiserum in Western blots. [18,21,26–28]

*Functionalization of Cantilevers*

To appropriately fix human fibronectin (Fn) to the cantilever tips, we employed a custom protocol to prepare our cantilevers. [29] Our functionalization scheme is illustrated in Figure 1(a). Before functionalization, fifteen cantilevers (circular symmetric and reflective gold coated for drift compensation) were loaded into a custom Teflon jig and incubated in chloroform (Sigma Aldrich) twice for ten minutes to remove organic contaminants. Subsequently, the same Teflon jig was then dipped in 20 mL 19:1 (v/v) piranha etch solution for 30 min. After etching, cantilevers were rinsed with deionized water, dried under a gentle argon stream, and re-incubated twice in chloroform for 10 min. Once the cantilevers were dried and washed, the entire jig was placed in a desiccator along with 30 µL of (3-aminopropyl) triethoxysilane (APTES) (Sigma Aldrich), 10 µL of triethylamine (TEA) (Sigma Aldrich), and allowed to incubate for two hours. TEA was included in the aminosilanization process to catalyze the reaction on the probe surface. [30] After two hours, the APTES and TEA were removed, and the desiccator was flushed with ultra-high purity argon gas. Cantilevers were allowed to be cured under argon for two days before further modification.[29,31,32]

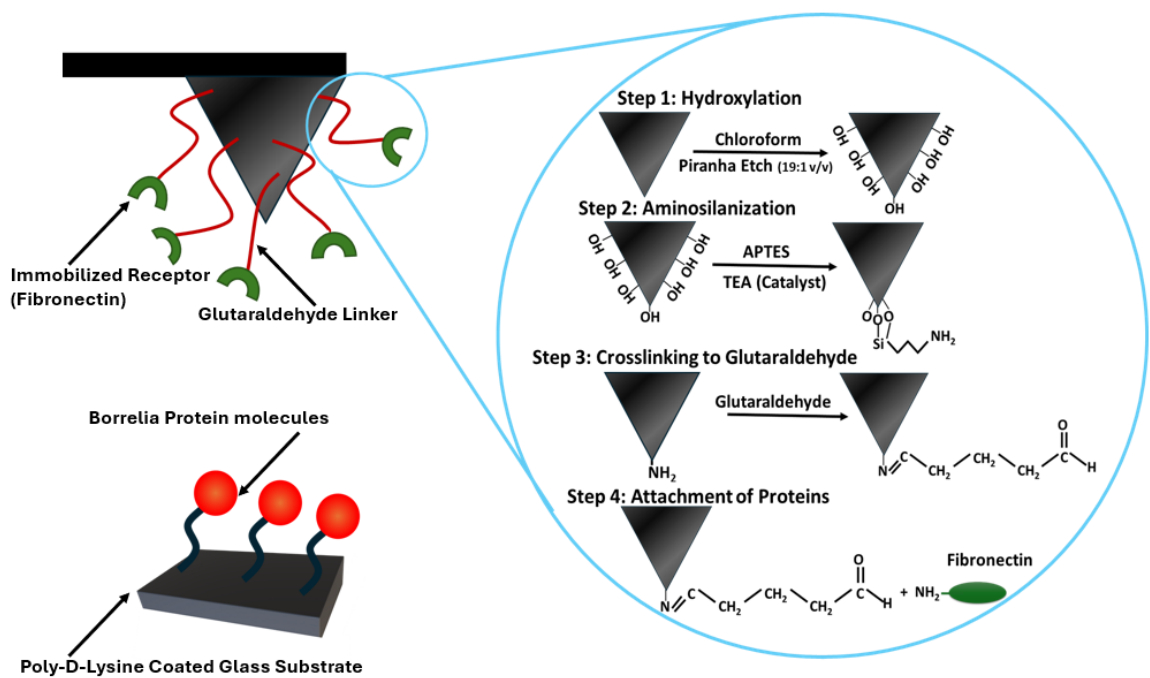

*Figure 1:* Coupling a functionalized AFM tip to Fn through a glutaraldehyde linker attachment strategy alongside the immobilization of Borrelia surface proteins onto a glass substrate.

Once the cantilevers were aminosilanized, they were functionalized with human Fn (Sigma Aldrich). First, a glutaraldehyde crosslinker (Sigma Aldrich) was attached to the free amine end of any adsorbed APTES molecules on the probe by incubating the cantilevers in a 1% (w/v) glutaraldehyde solution for 30 min. After incubation, the cantilevers were removed and rinsed extensively in deionized water to remove any free glutaraldehyde. [31] To functionalize with Fn, the as-prepared cantilevers were incubated in a 20mL phosphate buffer saline solution (PBS) with Fn at a concentration of 10μg Fn/ml PBS for 30 minutes. Functionalized cantilevers were stored at 4 ºC in PBS and remade as necessary.

*Single-molecule Force Spectroscopy Experiments*

After incubating proteins in a PDL-coated petri dish (FD35PDL, FluoroDish, WPI), we rinsed the dishes twice with 2 mL of fresh PBS (pH 7.4) to remove any non-adhered or weakly adhered proteins. Immediately following the rinsing, we added 2 mL of fresh buffer to prevent the sample substrate from drying out. Once the petri dish was placed on the sample stage, the functionalized cantilever was immersed in the solution and allowed to thermally acclimate for twenty minutes before taking measurements. The force-distance (FD) curves were recorded using a Bruker NanoWizard-4XP AFM, which was integrated with an inverted optical microscope (Olympus IX73) and equipped with top-view optics at $12 \times$ magnification. The AFM was situated in an anechoic chamber on an actively damped vibration isolation table to minimize noise during scanning. For the FD measurements, the qp-BioAC cantilever (NanoAndMore USA Corp.) was employed, featuring a nominal spring constant of 60 pN/nm and a frequency of 30 kHz. The cantilevers were calibrated in liquid using the thermal noise method. Throughout all experiments, the cantilevers demonstrated an average spring constant of $50.88 \pm 8.47$ pN/nm and an average resonance frequency of $7.07 \pm 0.71$ kHz.

FD measurements were performed at room temperature with a setpoint of 600 pN, a z-length of 500 nm, and no contact time. FD curves were acquired by recording at least 500 curves for six different pulling speeds, from 0.2 to 10 $\mu m/s$, on each *B. burgdorferi* surface protein. At least five independent experiments were conducted for each protein.

Control experiments were conducted to confirm the specificity of fibronectin interactions with bacterial surface proteins. The functionalized cantilevers were tested against a substrate coated with BSA (66.5 kDa, 583 aa) as opposed to bacterial surface proteins. BSA was selected due to its non-adhesive properties, serving as a blocking agent for the active sites of the molecules. [33] We conducted another control experiment using a surface coated with FlaB, a flagellin protein situated between the inner and outer membranes of the bacterium, which is crucial for the bacterium's motility and structural integrity. [34]

The recorded FD curves were filtered using the built-in software, JPK SPM, to eliminate non-specific binding events. In this step, we discarded FD curves that exhibited no adhesion, indicated non-specific interactions, or displayed multiple binding events, as noted in Ref. [35]. Typically, the rupture event is associated with a nonlinear elongation force, which essentially reflects the stretching of a flexible linker. Examples of the discarded representative FD curves can be found in Supplementary Figure S1. A one-way ANOVA was used to evaluate the null hypothesis of equal means across datasets ($p \geq 0.05$ indicating consistency). If the ANOVA was significant ($p < 0.05$), we conducted Tukey-Kramer post-hoc comparisons to identify and exclude datasets whose means differed significantly. The distribution of rupture forces among the groups is illustrated in the boxplot figures provided in the supplementary materials, accompanied by the corresponding ANOVA test results. A minimum of three independent groups was utilized to investigate the molecular dynamics of specific bonds further, employing the Bell-Evans model to derive the relevant parameters.

**Results & Discussion**

We investigated how *B. burgdorferi* surface proteins bind to Fn using an atomic force microscopy (AFM)-based single-molecule spectroscopy (SMFS) approach. Fn-conjugated AFM tips were brought into contact with a surface coated in surface proteins—RevA, BBK32, BmpA, OspA, and OspC—while FlaB (a flagellin protein) and BSA were used as controls. The tips were then pulled at varying speeds of 0.2, 0.5, 1.0, 2.0, 5.0, and 10 $\mu m/s$.

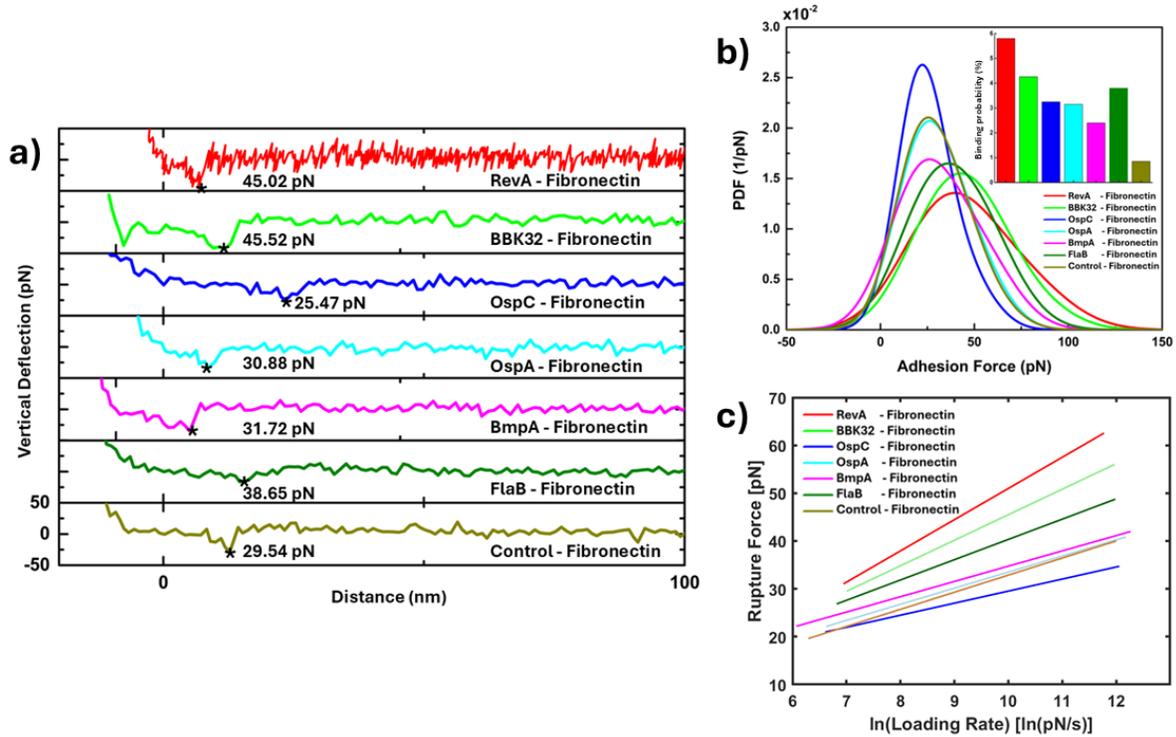

*Figure 2. a) Representative force-distance (FD) curves for each protein at a pulling speed of 2.0 µm/s. b) Probability density function plots illustrating the force distributions. The inset displays the binding probabilities, calculated as the percentage of FD curves that show specific binding events out of the total FD curves performed; control experiments with BSA indicated less than 1% binding. c) Rupture force plotted against loading rate for each protein. The orthogonal distance regression approach was employed*

*to fit the data, and the Bell-Evans model was utilized to extract the binding dynamics (original data and fits are provided separately for each protein in Supplementary Figures S4).*

Figure 2(a) presents the typical force-distance (FD) curves for each protein studied, highlighting average rupture forces obtained at a pulling speed of 2.0 µm/s. We estimated the most probable rupture forces by plotting the probability density functions (PDFs) and fitting them with a Gaussian function. [10] The $p(F)$ for a force distribution, estimated via the following equation:

$$p(F) = \sum_i \frac{1}{N\sigma\sqrt{2\pi}} \exp\left(\frac{-(F - \mu_i)^2}{2\sigma^2}\right)$$

Here, $p(F)$ is the estimated probability density at a chosen force $F$, $\mu_i$ is the experimentally measured rupture force for the $i^{th}$ event, and $\sigma$ is the global standard deviation for all successful events. We used a PDF plot instead of a histogram to obtain a smooth, binning-independent distribution of rupture forces, avoiding random fluctuations caused by histogram bin sizes. [36,37]

Figure 2(c) illustrates the PDF plots for each protein and the controls taken at a pulling speed of 2.0 µm/s, revealing that the binding strengths of RevA and BBK32 to Fn were notably higher than the others. This trend also holds for other pulling speeds, as shown in the PDF plots for each speed in supplementary Figure S3. Additionally, the inset in Figure 2(c) exhibits the binding probabilities, demonstrating that RevA and BBK32 generally exhibit higher binding probabilities than the other proteins, further highlighting their enhanced binding capacity across varying pulling speeds.

We used dynamic force spectroscopy (DFS) to characterize the molecular dynamics of the bonds. In this approach, the most probable rupture forces were plotted against their respective loading rates (the product of pulling speed and the effective spring constant). We then applied the Bell–Evans model [38,39] to extract kinetic parameters. Under the assumption of a single, force-independent energy barrier in a thermally activated regime, the model relates the most probable rupture force ($F$) to the loading rate ($\gamma$) via:

$$F = \frac{k_B T}{X_\beta} \ln\left(\frac{\gamma \cdot X_\beta}{k_{off} k_B T}\right)$$

where $k_B$ is the Boltzmann constant, $T$ is the absolute temperature, $X_\beta$ (the "dissociation length") is the barrier width, and $k_{off}$ is the kinetic off-rate at zero force, which relates to the barrier height of the free-energy barrier ($\Delta G_{act}$) through a Kramers/Arrhenius expression, $k_{off} \sim e^{-\Delta G_{act}/(k_B T)}$. [40,41] By fitting the rupture forces measured at different loading rates, one can extract $k_{off}$ and $X_\beta$ and then calculate the average bond lifetime $\tau$ under zero force from $\tau = 1/k_{off}$. This approach offers a quantitative measure of the strength and stability of protein-ligand or protein-protein interactions under mechanical stress. The extracted kinetic parameters were compared in Figure 3 and summarized in Table 2. The $X_\beta$ values for RevA and BBK32 seen in Figure 3(a) were notably distinct from the others, suggesting that they had a narrower energy barrier; RevA's width was even half that of the other proteins. Meanwhile, RevA's smaller $k_{off}$ value indicates a higher barrier height and consequently much slower bond rupture with a bond lifetime of $\tau = 70.126\ s$, whereas BBK32 shows roughly one-third of the barrier height of RevA. By contrast, the remaining proteins displayed similar, larger barrier widths and higher $k_{off}$ values, suggesting weaker, non-specific interactions. Consistent with their high rupture forces, RevA and BBK32 exhibit longer bond lifetimes, underscoring their enhanced mechanical stability when bound to fibronectin.

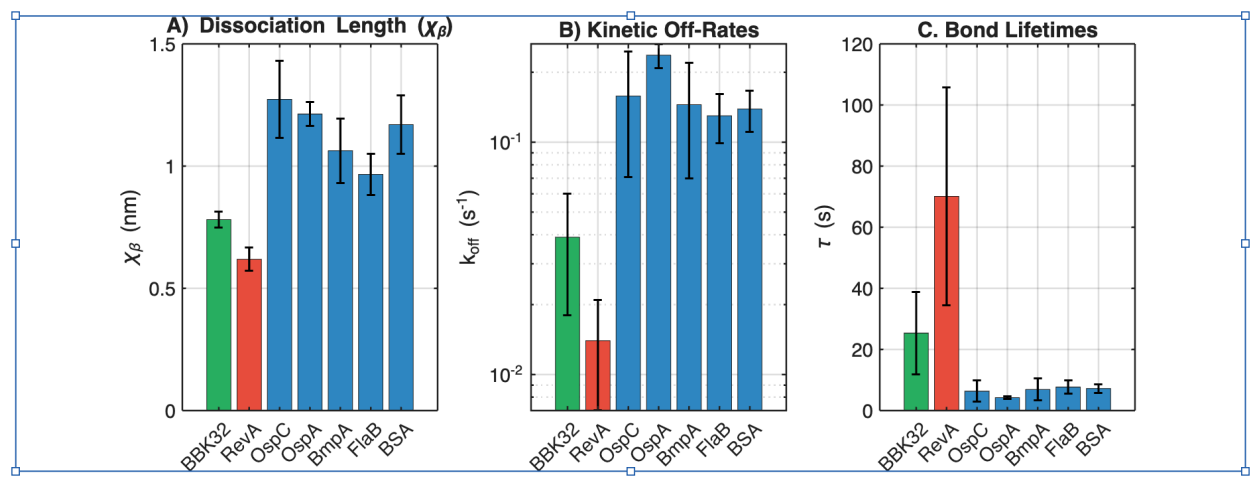

**Figure 3.** Comparison of interaction dynamics of *B. burgdorferi* adhesins with Fn. **a)** dissociation lengths, **b)** kinetic off-rates, and **c)** bond lifetimes.

**Table 2.** The dissociation path, $X_\beta$, the kinetic off-rate at zero applied force, $k_{off}$, and the average bond lifetime, $\tau$, under zero force for each protein.

| Protein Name | $X_\beta$ (nm) | $k_{off}$ (sec$^{-1}$) | $\tau$ (sec) |
|---|---|---|---|
| **BBK32** | 0.781 ± 0.033 | 0.039 ± 0.021 | 25.350 ± 13.400 |
| **RevA** | 0.619 ± 0.048 | 0.014 ± 0.007 | 70.126 ± 35.571 |
| **OspC** | 1.273 ± 0.157 | 0.158 ± 0.087 | 6.330 ± 3.500 |
| **OspA** | 1.213 ± 0.048 | 0.237 ± 0.028 | 4.224 ± 0.500 |
| **BmpA** | 1.063 ± 0.131 | 0.145 ± 0.075 | 6.903 ± 3.582 |
| **FlaB** | 0.966 ± 0.084 | 0.130 ± 0.031 | 7.690 ± 2.190 |
| **Control (BSA)** | 1.170 ± 0.120 | 0.139 ± 0.028 | 7.210 ± 1.440 |

Our AFM-SMFS measurements show that, across various pulling speeds, RevA and BBK32 exhibit higher rupture-force peaks and binding probabilities to fibronectin than OspA, OspC, or BmpA, indicating stronger interactions under our assay conditions. Using the Bell-Evans model, these trends are accompanied by smaller dissociation lengths ($\chi_\beta$) and lower zero-force off-rates ($k_{off}$), especially for RevA, resulting in longer lifetimes ($\tau$) and consistent with a narrower, deeper energy barrier under load. These kinetic signatures agree with previous reports identifying RevA and BBK32 as bona fide fibronectin-binding adhesins in *B. burgdorferi*. [18,28,42] In contrast, the weaker force readouts for OspA/OspC/BmpA align with their documented roles, OspC's broader, context-dependent functions, and BmpA's preference for laminin, rather than primary fibronectin anchoring. [43,44] Together, the rupture-force distributions and Bell–Evans parameters (Table 2) support partially non-redundant contributions of RevA/BBK32 to early, force-bearing fibronectin attachment. Finally, since these interactions initiate at the host interface, our findings suggest that targeted follow-up assays (e.g., competition with fibronectin fragments or blocking reagents) could be used to test whether reducing RevA/BBK32–fibronectin binding can decrease early attachment.

**Conclusion**

This study employed AFM-based single-molecule force spectroscopy and Bell–Evans dynamic force spectroscopy to systematically compare the engagement of *B. burgdorferi* adhesins with fibronectin under mechanical load. The findings indicate that RevA and BBK32 exhibit significantly stronger and more stable interactions with fibronectin relative to other adhesins, suggesting their distinct roles in mediating tissue colonization. Conversely, the remaining adhesins demonstrated larger barrier widths and elevated kinetic off-rate constants, signaling weaker or more nonspecific interactions. Collectively, these results elucidate a quantitative mechanism by which *B. burgdorferi* surface lipoproteins interact with fibronectin, highlighting the force resilience of RevA and BBK32. Importantly, by identifying these robust interactions, this work lays the foundation for targeted interventions aimed at disrupting early host-surface attachment, potentially informing strategies to mitigate infection.

**Author Contributions:**

K.M.H: Performed experiments, analyzed data, wrote the paper

C.M.: Performed experiments, analyzed data, wrote the paper

M.O.: Performed experiments, analyzed data, wrote the paper

C.A.B: Designed experiments, interpreted data, wrote the paper

N.O: Designed experiments, interpreted data, wrote the paper

**Declaration Statement:**

The authors declare no competing interests.

**Acknowledgement:**

We thank Dr. David Pierce for providing safety training on handling strong acids to K.M.H. and C.M., and for generously offering lab space to ensure the safe preparation and use of these chemicals.

The research presented in this paper was supported by the National Science Foundation under NSF EPSCoR Track-1 Cooperative Agreement OIA #1946202 and NSF-MRI Grant No. 2117014. Any opinions, findings, and conclusions or recommendations expressed in this material are those of the author(s) and do not necessarily reflect the views of the National Science Foundation.